\documentstyle[12pt]{article}
\newcommand{\ra}{\rightarrow}

\newcommand{\be}{\begin{equation}}
\newcommand{\ee}{\end{equation}}
\newcommand{\ba}{\begin{eqnarray}}
\newcommand{\ea}{\end{eqnarray}}

\newcommand{\dta}{\mbox{$\delta$}}

\newtheorem{prop}{Proposition}[section]

\newtheorem{rmk}{Remark}[section]

\begin{document}

\title{Modeling Vocal Fold Motion with a \\Continuum 
Fluid Dynamic Model:\\ I. Derivation and Analysis.}
\author{J. Xin \thanks{
Department of Mathematics and TICAM, Univ of Texas,
Austin, TX 78712.
}, J. M. Hyman \thanks{Mail Stop B284, 
Los Alamos National Lab, Los Alamos, NM 87545.  
}, 
and Y-Y Qi \thanks{Qualcomm Inc, 5775 Morehouse Drive, San Diego, CA 92121.
} }
\date{}
\maketitle
\date{}

\begin{abstract}
Vocal fold (VF) motion is fundamental to voice production and diagnosis in 
speech and health sciences. The motion is a consequence of air flow 
interacting with elastic vocal fold structures. 
Motivated by existing lumped mass models and known flow properties,  
we propose to model the continuous shape of vocal fold in motion by 
the two dimensional compressible Navier-Stokes equations coupled with
an elastic damped driven wave equation on the fold cover. 
In this paper, instead of pursuing a direct two dimensional 
numerical simulation, 
we derive reduced quasi-one-dimensional model equations by 
averaging two dimensional solutions along the flow cross sections. 
We then analyze the oscillation modes of the 
linearized system about a flat fold, 
and found that the fold motion goes through a Hopf bifurcation into 
temporal oscillation if the flow energy is sufficient to overcome 
the damping in the fold consistent with the early models. 
We also analyze the further reduced system under the quasi-steady
approximation and compare the resulting vocal fold equation in the small
vibration regime with that of the Titze model. Our model shares several
qualitative features with the Titze
model yet differs in the specific form of 
energy input from the air flow to the fold. 
Numerical issues and results of the quasi-one-dimensional 
model system will be presented in part II 
(view resulting web VF animation at http://www.ma.utexas.edu/users/jxin). 

\end{abstract}

\baselineskip=17.74pt

\thispagestyle{empty} \newpage \setcounter{page}{1}

\section{Introduction}
\setcounter{equation}{0}

Vocal folds consist of two lips of opposing ligaments and muscle at the  
top of the trachea and join to the lower vocal tract. When air is expelled
at sufficient velocity through this orifice (the glottis) from the lung 
after a critical lung pressure is achieved, 
the folds vibrate and act as oscillating valves interrupting the 
airflow into a series of pulses. These pulses of air flow serve as 
the excitation source for the vocal tract in all voiced sounds. 
For background description of vocal folds and speech production, see 
\cite{Flan1}, \cite{Fry}, \cite{Ti0} among others.  
Mathematical modeling of vocal fold motion can help us understand 
the voiced sound generation and make synthetic machine 
voice more natural. Such an understanding also helps the 
medical diagnosis and correction of voice disorders \cite{Ishi}, \cite{Merg}. 

Vocal fold modeling relies on our knowledge of aerodynamics and 
biomechanics, as the problem is basically airflow through a flexible channel 
with time-varying cross section. 
One of the best known models of vocal folds is the 
two mass model by Ishizaka and Flanagan \cite{Ishi}. The vocal fold is 
modeled by upper and lower masses ($m_i$, $i=1,2$) 
connected by an elastic spring 
and attached to the wall by upper and lower springs with damping. 
The two mass motion is given by the ODE system:
\be
m_{i}x_{i,tt} + r_{i}x_{i,t} + k_{i}x_{i} + k_{c}(-1)^{i+1}(x_1 -x_2) = F_i, 
\;\; i=1,2, \label{I1}
\ee
where $r_i$'s are the damping coefficients; $k_i$'s, $k_c$, are the 
elastic restoring constants (stiffness coefficients) for the upper, lower 
and side springs respectively; $x_i$'s are the displacements of two 
masses from their prephonatory equilibrium positions; $F_i$'s are driving 
forces acting on the two masses from glottal air pressure. To close the 
equations and complete the model, simplified assumptions are made to 
allow a calculation of $F_i$ from the fluid flow so that 
$F_i =F_i(x_1,x_2,d_i, l_i, P_s)$, where $P_s$ is the lung pressure, 
$d_i$ the thickness of 
$m_i$ along the flow direction, $l_i$ the length of $m_i$ transverse to 
the flow direction. The flow is assumed to be one dimensional, inviscid, and 
quasisteady so that Bernoulli's law can be applied. Moreover, $F_i$ and 
elastic forces $k_i x_i$ need to be modified empirically when the vocal 
folds tend to a closure to facilitate a pressure buildup to reopen the 
folds. In spite of these oversimplifications, the two mass model is 
able to capture many aspects of oscillation features. Small 
amplitude oscillation 
appears as a Hopf bifurcation from a steady state as $P_s$ passes a 
threshold value \cite{Ishi1}. More recently, 
Lucero \cite{Luc0} studied large amplitude oscillations and 
found coexistence of 
multiple equilibria and studied their stability and bifurcations. 
 
The two mass model has been widely used as a glottal source for 
speech synthesis, and many improvements and extensions have been 
made to incorporate additional physics. For example, see \cite{Bo}
for including effects of flow viscosity and flow separation; see \cite{Smith}, 
\cite{Wong}, for works on four and multiple mass models and applications.  
In \cite{Berry}, a continuum elastic system under natural 
boundary condition is used to find intrinsic vibrational 
eigenmodes of vocal fold tissues.  

Though two mass model is already simple looking, the theshold oscillation 
condition on minimum lung pressure is still implicit analytically. 
Titze proposed his celebrated body-cover model in 1988 \cite{Ti}. 
The model is based on the hypothesis that during fold oscillation, 
the fold cover (epithelium and superficial layers of vocal ligament) 
propagates a surface mucosal wave in the direction of the airflow, and 
the body (deep layer of the vocal ligament and muscle) is stationary. 
The fold shape is
approximated as a straight line connecting the fold entry of height $h_{1}$
and the fold exit of height $h_{2}$. Taylor expanding a mucosal wave with
constant velocity $c$, Titze \cite{Ti} approximated:
\begin{equation}
h_{1}=2(h_{0}+\hat{h}+\tau \hat{h}_{t}),\;h_{2}=2(h_{0}+\hat{h}-\tau 
\hat{h}_{t}),  \label{I2}
\end{equation}
where $\tau =2L/c$ the travel time for the mucosal wave to reach exit from
entrance, $L$ half length of glottis along the flow direction, 
$h_0$ half glottal width, $\hat{h}$ fold displacement. 
The fold motion is lumped onto the fold midpoint, and postulated as: 
\begin{equation}
m\hat{h}_{tt}+r \hat{h}_{t}+k \hat{h}=
{\frac{1}{2L}}\int_{-L}^{L}\,P(x)\,dx\equiv P_{g},  \label{I3}
\end{equation}
where $m$, $r$, and $k$ are the mass, damping, and stiffness of 
the oscillating portion of the vocal fold about the midpoint, 
$P(x)$ is the pressure distribution along the flow direction $x$.
Using Bernoulli's law and linear fold shape, \cite{Ti} showed in
particular that for small fold vibration about rectangular fold equilibrium
position $h_{0}$: 
\begin{equation}
P_{g}=P_{s}(1-{\frac{h_{2}}{h_{1}}})={\frac{2\tau P_{s}\hat{h}_{t}}{h_{0}+
\hat{h}+\tau \hat{h}_{t}}},  \label{I4}
\end{equation}
where $P_{s}$ is the subglottal pressure, and exit pressure is zero.
Linearizing (\ref{I4}) about $h_{0}$, one has: 
\begin{equation}
m\hat{h}_{tt}+(r -2\tau P_{s}h_{0}^{-1})\hat{h}_{t}+ k \hat{h}=0,
\label{I5}
\end{equation}
from which follows the threshold pressure condition 
setting the effective damping coefficient to zero: 
\begin{equation}
P_{s,*}= h_{0}r /2\tau.
\label{I6}
\end{equation}
So the oscillation threshold pressure is proportional to damping constant 
$r$, and glottal half width $h_0$.

Titze model explains in a transparent fashion the formation of 
 oscillation, and provides a very 
simple onset condition (\ref{I6}). It shows how the energy input from 
the airflow balances the intrinsic fold damping due to the  
$\hat{h}_{t}$ dependence of the fold driving force $P_g$. 
The agreement of Titze model and the two mass model is discussed in 
\cite{Ti}, \cite{Ti0}. The onset condition (\ref{I6}), especially 
$P_{s,*}$ being linear in $r$, is also supported 
by experiments \cite{Ti1}. For results of Titze model on converging 
and diverging prephonatory fold shapes, large amplitude 
oscillations, and hysteresis phenomenon at onset-offset, see 
\cite{Ti1}, \cite{Luc1}, \cite{Luc2}. 

In either the two mass model or Titze model, 
the vocal fold shape is not described as a continuous curve as we 
see in physical reality, and 
that the treatment of air flow is rather crude. It is our goal  
to seek a more accurate and more first principle based model so that 
these two aspects are much improved. The fluid characteristics of 
vocal flow during phonation have been noted in  
\cite{Ishi}, \cite{Ti}, \cite{Ti0}, \cite{Luc0}, \cite{Luc2}, 
among others, and also recently studied by Pelorson 
et al \cite{Pel1} using unsteady-flow measurements and visualization
techniques. The flow is essentially two dimensional with Mach number 
on the order of $10^{-1}$, and Reynolds number on the order of $10^{3}$. 
The flow in the bulk is nearly inviscid except in the viscous boundary 
layers. 
    
Our starting point is the two dimensional 
compressible isentropic Navier-Stokes equations for the air 
flow. Motivated by \cite{Ishi} and \cite{Ti}, 
we model the vocal fold as a finite mass elastic tube of cross sectional
area $A(x,t)$ with elastic attachments (muscles) onto nearby walls (bones). 
We shall either take rectangular cross section (length $2L
$ in $x$, width $2w$ in $z$, and variable height $2h$ in $y$, 
then $A = 4 w h$) or 
an elliptical shaped cross section (with principal axes 
of lengths $2h$ in $y$, and 
$2w$ in $z$). The air flows from $x= -L$ to $x=L$, symmetric across 
$y=0$, and is independent of $z$. 
The purpose of keeping the $z$ dimension is for later reconstruction of 
a three dimensional picture of fold motion in part II \cite{Mac}.  
Assuming that the flow is essentially along $x$ direction, and its 
variation transverse to the tube is small except in the boundary layers 
near the folds, 
we average the flow on each cross section and derive a reduced 
quasi-one-dimensional system. The motion of fold cover is modeled by 
a damped driven elastic wave equation, considering the 
elastic tension on the fold cover and elastic forces in the attached muscles. 

The reduced quasi-one-dimensional aerodynamic equations are:

\noindent $\bullet$ conservation of mass: 
\begin{equation}
( A\rho )_{t} + (\rho\, u \, A)_{x} =0,  \label{E1}
\end{equation}
$\rho$ air density, $u$ air velocity;

\noindent $\bullet$ conservation of momentum: 
\begin{equation}
u_{t} + u u_{x} = - {\frac{1}{\rho}}p_{x}+ {\frac{A_{t}\, u}{A}} 
+ {\frac{4\mu }{3\rho}} A^{-1}(A \, u_{x})_{x} 
- {2\mu \over 3\rho } A^{-1}\,A_{tx},  \label{E2}
\end{equation}
$p$ air pressure, $\mu$ air viscosity.

\noindent The force balance on the fold cover gives:

\noindent $\bullet$ the dynamic boundary motion: 
\begin{equation}
m(A-A_{eq})_{tt} = \sigma (A-A_{eq})_{xx} - \alpha (A-A_{eq})_{t} - \beta
(A-A_{eq}) + S p + f_m.  \label{E3a}
\end{equation}

\noindent Here: $m$ is the fold mass density; $\sigma$ is the longitudinal
elastic tension of the fold \cite{Fung}, \cite{Rossing}; 
$\alpha$ is the muscle damping coefficient; $\beta$
is an elastic modulus modeling the vibration property of the fold in the
transverse; $S$ is a cross section shape factor, $S=4w$ for
rectangular cross sections, and $S=\pi w$ for elliptical cross sections; $f_m
$ is a prescibed function to model muscle tone so that a particular fold
shape $A_{eq}$ (converging or diverging or flat) serves as an equilibrium
state. In general, $\alpha$, $\beta$, $\sigma$ are functions of $x$, 
to model the varying stiffness of fold, since the fold cover is stiffer 
towards $x=\pm L$. 

We shall write (\ref{E3a}) into: 
\begin{equation}
m A_{tt}= \sigma A_{xx} -\alpha A_{t} -\beta A + S p + \tilde{f}_{m},
\label{E3}
\end{equation}
where $\tilde{f}_{m}=\tilde{f}_{m}(x)$ is prescribed forcing. 

\noindent $\bullet$ The equation of state: 
\begin{equation}
p = \kappa \, \rho^{\mbox{$\gamma$}}, \; \mbox{$\gamma$} > 1,\; \kappa > 0.
\label{E4}
\end{equation}

The system (\ref{E1})-(\ref{E4}) is closed, and we solve an initial boundary
value problem on $x \in [-L,L]$ with proper in-flow/out-flow 
boundary conditions and initial fold shape. 
We shall consider 
{\it the inviscid limit of our model} by setting $\mu = 0$ and normalize 
$m=1$ while analyzing the oscillation onset in this paper. 
We shall show via a 
stability analysis that the linearized system near a flat fold  
admits (fold cover) waves in $A$ of the form $\sin (kx -\omega t)$, 
for $k$ and $\omega$ real and nonzero if the glottal airflow has 
enough energy to overcome fold damping. This is the kind of fold surface 
mucosal wave hypothesized in Titze model \cite{Ti}. We also obtain 
similar onset conditions for small amplitude oscillations due to 
the occurence of a pair of imaginary eigenvalues, consistent with 
early models. Under the quasi-steady approximation as in Titze model, 
we obtain a single fold equation resembling (\ref{I5}), the correction 
term is nonlocal but does depend on $\hat{h}_{t}$, and it plays the role of 
transferring aerodynamic energy onto the fold. 

We remark that the viscous effect is important when the fold is 
near closure, or when the fold is diverging enough so that 
flow separation occurs inside the glottis \cite{Ti}, \cite{Pel1}, \cite{Bo}. 
Our model above has not yet taken this aspect into full consideration. 
However, since flow separation decreases the pressure 
(pressure is zero from the point where flow detaches from the fold ) 
towards the exit $x=L$, and the fold is relatively stiff there, 
we can minimize the 
effect on fold motion by modeling the increasing stiffness of fold near 
the exit. Our model captures the compressibility of air flow which 
is critical during the opening of the folds \cite{Mong}. The viscous 
effect during the opening or near closure of vocal fold 
is delicate and requires 
further modeling, yet we have not found the lack of it to be a major problem 
in our simulation of the opening and closing of fold motion in part II 
\cite{Mac}.     
       
In studying collapsible tubes (\cite{Shapiro1}, \cite{Ku}) and air flow
through duct of spatially varying cross section \cite{Whith}, it is common
to use (\ref{E1}) with the one dimensional unsteady 
Euler equation which is (\ref{E2}) with the last three terms omitted. 
The major differences in the two
modeling problems are: (1) a vocal fold is fast oscillatory in time (e.g.
100 - 200 Hz), (2) the vocal fold carries mass, and a {\it dynamic (damped
driven wave) equation } is necessary to describe the fold motion; moreover,
the vocal fold has mechanical damping. In contrast, collapsible tubes are
massless and damping free \cite{Shapiro1}. The conservation of mass and
momentum in the collapsible tube system is recovered when $A_t \ll A$ and $
\mu \to 0$. Though in both collapsible tube and vocal flow problems, 
the flow is close to being incompressible, we opt not to make such 
an approximation as in \cite{Shapiro1}. This is because an equation of 
state has to be determined in lieu of the natural equation of state 
(\ref{E4}) when we consider quasi-one-dimensional reduction. 
For collapsible tubes, 
see \cite{Shapiro1} and \cite{Ku}, the tube cross section is related to the
pressure $p$ by a tube law: $p = A^{n_1} -
A^{n_2}$ with $\rho$ = constant. The tube law needs experimental measurements 
and depends also on materials involved. Due to lack of our knowledge 
of whether such a law exists for vocal fold, we choose to use the 
natural equation of state for air (\ref{E4}) and work with compressible 
flow equations. It is interesting to investigate how good it is to use the 
incompressible Navier-Stokes equations for the two dimensional flow 
as a viable alternative. To rephrase our formulation, the 
$\rho$ and $p$ are related by the equation
of state, the gamma gas law (\ref{E4}); 
then the cross section $A$ is related to $p$ {\it dynamically}.

The rest of the paper is organized as follows. In section 2, we derive 
the fluid part of the 
quasi-one-dimensional system (\ref{E1})-(\ref{E4}) from the 
two dimensional isentropic compressible Navier-Stokes equations by 
averaging solutions over the flow cross section. In section 3, we 
perform a linear stability analysis near a flat (rectangular) fold and 
deduce onset condition for the minimum flow energy to ensure the 
occurence of a pair of pure imaginary eigenmodes. In section 4, we 
make the quasi-steady approximation within our model, derive the 
single equation for the fold cover, obtain an onset condition for 
small amplitude oscillation, 
and compare with findings in \cite{Ti}. 
Numerical issues, and numerical results 
will be presented in part II \cite{Mac}.        
 
\section{Derivation of Reduced Flow Equations}
\setcounter{equation}{0}

We derive the fluid part of the model system assuming that the fold varies
in space and time as $A=A(x,t)$. Consider a two dimensional slightly viscous
subsonic air flow in a channel with spatially temporally varying cross
section in two space dimensions, $\Omega_0 =\Omega_0(t) = \{ (x,y): x \in
[-L,L], y \in [-A(x,t)/2,A(x,t)/2]\}$, where $A(x,t)$ denotes the channel
width with a slight abuse of notation, or cross sectional area since
 the third dimension is uniform. The two dimensional Navier-Stokes
equations in differential form are (\cite{Batch}, page 147):

\noindent $\bullet$ conservation of mass: 
\begin{equation}
\rho_{t} + \nabla \cdot (\rho\, \vec{u}) =0;  \label{E5}
\end{equation}

\noindent $\bullet$ conservation of momentum: 
\begin{equation}
(\rho \vec{u})_t = - \nabla \cdot (\rho\, (\vec{u}\otimes \vec{u})) +
div ( \sigma );  \label{E6}
\end{equation}
where $\sigma$ is the stress tensor, $\sigma =(\sigma_{ij})= - p %
\mbox{$\delta$}_{ij} + d_{ij}$, and: 
\[
d_{ij}= 2\mu\, (e_{ij} - {\frac{div \vec{u} }{3}}\mbox{$\delta$}_{ij}),\;\;
e_{ij} = {\frac{1}{2}}(u_{i,x_{j}} + u_{j,x_{i}}), \;\; (x_1,x_2) \equiv
(x,y); 
\]
$\mu$ is the fluid viscosity; $\Omega (t)$ is any volume 
element of the form: 
\be
\Omega (t)= \{ (x,y): x \in [a,b] \subset [-L,L], y \in [-A(x,t)/2,A(x,t)/2].
\}. \label{dom} 
\ee
The equation of state is (\ref{E4}).

The boundary conditions on $(\rho,\vec{u})$ are:

\noindent (1) on the upper and lower boundaries $y = \pm A(x,t)/2$, $%
\rho_{y} = 0$, and $\vec{u}=(0,\pm A_{t}/2)$, the velocity no slip boundary
condition;

\noindent (2) at the inlet, $\vec{u}(-L,y,t)= \vec{u}_{l}$, a prescribed
inlet velocity, $\rho(-L,y,t)=\rho_{l}$, a prescribed inlet density (deduced
from input pressure);

\noindent (3) at the outlet, $\vec{u}_{x}(L,y,t)=0$. 

We are only concerned with flows that are symmetric in the vertical. For
positive but small viscosity, the flows are laminar in the interior of $
\Omega_0$ and form viscous boundary layers near the upper and lower edges.
The vertically averaged flow quantities are expected to be much less
influenced by the boundary layer behavior as long as $A(x,t)$ is much larger 
than $O(\mu^{1/2})$. We also ignore effects of possible flow seperation
inside $\Omega_0$ when it becomes divergent with large enough opening. 

Let us assume that the flow variables obey: 
\begin{eqnarray}
|u_{1,y}| \ll |u_{1,x}|, \; |u_{2,y}| \ll |u_{1,x}|, \; {\rm away\; from \;
boundaries\; of\;} \Omega_0,  \nonumber \\
|\vec{u}_{y}| \gg |\vec{u}_{x}|, \; {\rm near\; the\; boundaries\; of\;}
\Omega_0,  \nonumber \\
|\rho_y| \ll |\rho_x|, \; {\rm throughout\;} \Omega_0.  \label{E7}
\end{eqnarray}
These are consistent with physical observations in the viscous boundary
layers (\cite{Batch}, page 302), 
namely, there are large vertical velocity gradients,
yet small pressure or density gradients in the boundary layers. 
The boundary layers are of width $O(\mu^{1/2})$. 
Denote by $\overline{\rho}$, $\overline{u%
}_{1}$, the vertical averages of $\rho$ and $u_1$. Note that the exterior
normal $\vec{n} = (-A_x/2,1)/(1+A_x^2/4)^{1/2}$ if $y=A/2$, $\vec{n} =
(-A_x/2,-1)/(1+A_x^2/4)^{1/2}$ if $y=- A/2$.

\noindent Let $a=x$, $b=x+\mbox{$\delta$} x$, $\mbox{$\delta$} x \ll 1$, 
$t$ slightly larger than $t_0$. We have:
\begin{equation}
{\frac{d}{dt}}\int_{\Omega (t)} \, \rho\, dV = {\frac{d}{dt}}\int_{\Omega
(t_0)} \, \rho\, J(t)\, dV = \int_{\Omega (t_0)}\, \rho_t \, J(t)\, dV +
\int_{\Omega (t_0)}\, \rho\, J_t\, dV,  \label{J1}
\end{equation}
where $J(t)$ is the Jacobian of volume change from a reference time $t_0$ to 
$t$. Since $\Omega(t)$ is now a thin slice, $J(t) ={\frac{A(t) }{A(t_0)}}$
for small $\mbox{$\delta$} x$, and $J_t = A_{t}(t)/A(t_0)$. The second
integral in (\ref{J1}) is: 
\begin{equation}
\int_{\Omega (t_0)}\, \rho\, J_t\, dV = \overline{ \rho} 
{\frac{A_{t}(t)}{A(t_0)}}
A(t_0) \mbox{$\delta$} x = \overline{\rho} \, A_{t}(t) \mbox{$\delta$} x.
\label{J2}
\end{equation}
The first integral is simplified using (\ref{E5}) as: 
\begin{equation}
\int_{\Omega (t_0)}\, \rho_t \, J(t)\, dV = \int_{\Omega (t)} \, \rho_t\, dV
= - \int_{\partial \Omega (t)} \rho \, \vec{u}\cdot \vec{n}\, dS.  \label{J3}
\end{equation}
We calculate the last integral of (\ref{J3}) further as follows: 
\begin{eqnarray}
\int_{\partial \Omega} \rho \vec{u}\cdot \vec{n} \, ds 
& = & \int_{-A/2}^{A/2}\,
(-\rho u_{1})(x,y,t) \, dy + \int_{-A/2}^{A/2}\, (\rho u_{1})(x+%
\mbox{$\delta$} x,y,t) \, dy  \nonumber \\
& + & \int_{x}^{x+\mbox{$\delta$} x}\, \rho\cdot (0,A_t/2)\cdot
(-A_{x}/2,1)\, dx  \nonumber \\
&+ & \int_{x}^{x+\mbox{$\delta$} x}\, \rho\cdot (0,-A_t/2)\cdot
(-A_{x}/2,-1)\, dx  \nonumber \\
& = & \overline{\rho u_{1}} A|^{x+\mbox{$\delta$} x}_{x} + {\frac{%
\mbox{$\delta$} x}{2}}(\rho\, A_{t})|_{y=A/2} + {\frac{\mbox{$\delta$} x}{2}}%
(\rho\, A_{t})|_{y=-A/2}+ O((\mbox{$\delta$} x)^2)  \nonumber \\
& \approx & (\overline{\rho}\cdot \overline{u}_{1} A)|^{x+\mbox{$\delta$}
x}_{x}+ \overline{\rho}A_{t} \mbox{$\delta$} x + O((\mbox{$\delta$} x)^2),
\label{E9}
\end{eqnarray}
where we have used the smallness of $\rho_{y}$ to approximate $\rho|_{y=\pm
A/2}$ by $\overline{\rho}$ and $\overline{\rho u_{1}}$ by $\overline{\rho}%
\cdot \overline{u_{1}}$.

\noindent Combining (\ref{J1})-(\ref{J3}), (\ref{E9}) with: 
\begin{equation}
{\frac{d}{dt}} \int_{\Omega} \rho \, dV =-(\overline{\rho}\cdot \overline{u}_{1} A)|^{x+\mbox{$\delta$}x}_{x} 
 + O((\mbox{$\delta$} x)^2),  \label{E8}
\end{equation}
dividing by $\mbox{$\delta$} x$ and sending it to zero, we have: 
\[
(\overline{\rho}A)_{t} + (\overline{\rho}\cdot \overline{u}_{1} A)_{x}  = 0, 
\]
which is (\ref{E1}).

Next consider $i=1$ in the momentum equation, $a=x$, $b=x+\mbox{$\delta$} x$%
. We have similarly with (\ref{E6}): 
\begin{eqnarray}
{\frac{d }{dt}} \int_{\Omega (t)} \, \rho\, u_1\, dV  =  \int_{\Omega (t)}
(\rho u_{1})_{t}\, dV + \int_{\Omega (t_0)} \rho\, u_1\, J_t\, dV  \nonumber
\\
 = - \int_{\partial \Omega (t)}\rho u_{1} \vec{u}\cdot \vec{n}\, dS +
\int_{\partial \Omega (t)} \sigma_{1,j}\cdot \vec{n}_{j}\, dS 
 + \overline{\rho}\, \overline{u_1}\, A_{t} %
\, \mbox{$\delta$} x + O(\delta x^2).  \label{J4}
\end{eqnarray}

\noindent We calculate the integrals of (\ref{J4}) below. 
\be
{\frac{d }{dt}}\int_{\Omega } \rho u_{1}\, dV  = 
 (\overline{\rho}\overline{u}
_{1}A)_{t}\mbox{$\delta$} x + O((\mbox{$\delta$} x)^2) \approx (\overline{%
\rho}\cdot \overline{u} A)_{t}\cdot \mbox{$\delta$} x + O((\mbox{$\delta$}
x)^{2}), \label{key1}
\ee
where $u_1=0$ on the upper and lower boundaries is used. 
\be
\int_{\partial \Omega}\rho u_{1} \vec{u}\cdot \vec{n}\, dS = (\overline{\rho}%
\cdot \overline{u}_{1}^{2}A)|_{x}^{x+\mbox{$\delta$} x} + 
O(\dta x \, \mu^{1/2}),
\label{conv}
\end{equation}
where the smallness of $u_{1,y}$ in the interior and small width of boundary
layer $O(\mu^{1/2})$ gives the $O(\mu^{1/2})$ for approximating $\overline{%
u_{1}^2}$ by $\overline{u_1}\cdot \overline{u_{1}}$.

\begin{eqnarray}
\int_{\partial \Omega} - p \mbox{$\delta$}_{1,j}n_{j} dS &\approx & -%
\overline{p}A|_{x}^{x+\mbox{$\delta$} x} + \int_{x}^{x+\mbox{$\delta$} x}
p\, A_{x}\, dx  \nonumber \\
& = & -\overline{p}A|_{x}^{x+\mbox{$\delta$} x} + \overline{p}\, A_{x}\, %
\mbox{$\delta$} x + O( (\mbox{$\delta$} x)^{2}).  \nonumber
\end{eqnarray}

\noindent Noticing that:
\[
d_{11} = 2\mu (u_{1,x} - (u_{1,x} + u_{2,y})/3 ),\; 
d_{12} = 2\mu (u_{1,y}+u_{2,x}). 
\]
It follows that 
\[ \overline{
d_{11}} = {\frac{4}{3}}\mu \overline{u}_{1,x} - {2\mu A_{t}\over 3 A}.\]
Thus the
contribution from the left and right boundaries located at $x$ and $x +%
\mbox{$\delta$} x$ is: 
\begin{equation}
\sum_{l,r}\int_{l,r} \, d_{11}n_{1} =  A\, \overline{d_{11}}|_{x}^{x+%
\mbox{$\delta$} x} = {\frac{4}{3}}A \, \mu\, \overline{u}_{1,x}|_{x}^{x+%
\mbox{$\delta$} x} - {2\mu \, A_{t}\over 3}|^{x+\delta x}_{x}.
\label{E11}
\end{equation}
The contribution from the upper and lower boundaries is: 
\begin{eqnarray}
\sum_{\pm}\int_{y=\pm A/2} d_{11}\, n_{1}\, dS \, & = & - d_{11}A_{x}%
\mbox{$\delta$} x/2 |_{y=A/2} - d_{11}A_{x}\mbox{$\delta$} x/2 |_{y=-A/2} 
\nonumber \\
& = & \mu \mbox{$\delta$} x \sum _{\pm} O(\partial_{y} \vec{u})|_{y=\pm
A/2}.  \label{E12}
\end{eqnarray}
Similarly, 
\begin{equation}
\sum_{\pm}\int_{y=\pm A/2}d_{12} n_{2}\, dS = \mu \mbox{$\delta$} x
\sum_{\pm} O(\partial_{y}\vec{u})|_{y=\pm A/2}.  \label{E13}
\end{equation}

Since $\partial_{y} \vec{u}|_{y=\pm A/2} =O(\mu^{-1/2})$, the viscous
flux from the boundary layers are $O(\dta x \mu^{1/2})$,  much larger than the
averaged viscous term $\dta \, x\, 
{\frac{4\mu }{3}}(A\overline{u_{1}}_{x})_{x}= O(\dta \, x \, \mu)$.
We notice that the
vertically averaged quantities have little dependence on the viscous
boundary layers unless $A$ is on the order $O(\mu^{1/2})$. 
Hence the quantities from upper and lower edges in (\ref
{E12}) and (\ref{E13}), and that in (\ref{conv}), should balance themselves.
Omitting them altogether, and combining remaining terms that involve only $%
\overline{u_1}$, $\overline{\rho}$ in the bulk, we end up with 
(after dividing by $\delta x$ and sending it to zero): 
\begin{equation}
(\overline{\rho}\cdot \overline{u_1} A)_{t} +(\overline{\rho}\cdot \overline{%
u_1}^2 A)_{x} = -(\overline{p}\, A)_{x} + A_{x}\overline{p} 
+\overline{\rho}\, \overline{u_1}\, A_{t} 
+ {\frac{4\mu}{3}}
(A\overline{u_1}_{x})_{x} - 2\mu A_{tx}/3.  \label{E14}
\end{equation}
Simplifying (\ref{E14}) with the continuity equation (\ref{E1}), we find
 that: 
\begin{equation}
\overline{u_1}_{t} + \overline{u_1}\, \overline{u_1}_{x} = -\overline{p}_{x}/%
\overline{\rho} + {\frac{A_{t}\overline{u_1}}{A}} + {\frac{4\mu }{3\overline{%
\rho}}} A^{-1}(A \overline{u_1}_{x})_{x} - 
{2\mu A^{-1}\, A_{tx}\over 3 \overline{\rho}},  \label{E15}
\end{equation}
which is (\ref{E2}).

\section{Linear Stability Analysis near a Flat Fold}

\setcounter{equation}{0} 
In this section, we discuss the existence of a
neutral oscillation mode of the linearized system around a flat fold in the 
limit $\mu \ra 0$. 
Such a mode provides an onset condition of oscillation, 
see similar analysis for the Titze model in \cite{Ti} and \cite{Luc2}. It
turns out that such a mode exists under a theshold condition 
for system (\ref{E1})-(\ref{E4}) because of
the $A_t\, u/A$ term.  In our case, 
it is essential that the derivation originated with the no slip
boundary condition on the fold, which allows enough energy of the
background flow to transfer to the fold to offset the loss there. 
This energy transfer mechanism is expounded in Titze \cite{Ti} in his 
body-cover model. We show with a stability analysis 
that our model system supports such a physical 
picture of flow induced oscillation. 

We assume that the cross section is rectangular. 
The system (\ref{E1})-(\ref{E4}) admits constant steady states: $
(u_0,p_0,A_0)$, $p_0=\kappa \rho_0^{\mbox{$\gamma$}}$, satisfying: 
\begin{equation}
-\beta A_0 + 4 w p_0 = f_m,  \label{E35}
\end{equation}
where $f_m$ is a constant so that $A_0$ matches the height of the connecting
vocal tract. 
We are interested in conditions leading to the small amplitude oscillations
near the constant steady states. This is similar to Titze \cite{Ti}, where a
lumped ODE is proposed and analyzed for the fold center using mucosal wave
approximation. However here, we calculate directly from (\ref{E1}
)-(\ref{E4}), and do not make any further modeling assumptions.
Since we have zero background pressure gradient, our results do not
compare directly with \cite{Ti}, though qualitative features remain.
Another comparison with \cite{Ti} is performed in the next section 
under quasi-steady approximation where small pressure gradient is present. 

Letting $u=u_0 + \hat{u}$, $p= p_0 + \hat{p}$, $\rho=\rho_0 + \hat{\rho}$, $%
A = A_0 + \hat{A}$, and linearizing (\ref{E1})-(\ref{E4}) with $\mu=0$, 
we get: 
\begin{eqnarray}
(A_0\hat{\rho} + \hat{A}\rho_0)_{t} + (\rho_0 A_0\hat{u} + \rho_0 u_0 \hat{A}
+ u_0A_0\hat{\rho})_{x} = 0,  \label{E36} \\
\hat{u}_{t}+ u_{0}\hat{u}_{x} + {\frac{1}{\rho_0}}\hat{p}_{x} =
 {\frac{u_0}{A_0}}\hat{A}_{t},  \label{E37} \\
\hat{A}_{tt} = \sigma \hat{A}_{xx} - \mbox{$\alpha$} \hat{A}_{t} -\beta \hat{%
A} + 4w\hat{p}.  \label{E38}
\end{eqnarray}
Equations (\ref{E36})-(\ref{E37}) are written as: 
\begin{eqnarray}
A_0 \hat{\rho}_{t} +\rho_0\hat{A}_{t} + \rho_0 A_0 \hat{u}_{x} + \rho_0 u_0 
\hat{A}_{x} + u_{0}A_{0}\hat{\rho}_{x} = 0,  \label{E36a} \\
\hat{u}_{t} +u_{0}\hat{u}_{x} +{\frac{\hat{p}_{x}}{\rho_{0}}} =
 {\frac{u_0}{A_0}}\hat{A}_{t}.  \label{E37a}
\end{eqnarray}
Applying the operator $\partial_t +u_{0}\partial_x $ on (%
\ref{E36a}) and using (\ref{E37a}), we find: 
\begin{equation}
(\partial_t +u_{0}\partial_x )( A_0 \hat{\rho}_{t} + \rho_0%
\hat{A}_{t} + \rho_0 u_0 \hat{A}_{x} + u_{0}A_{0}\hat{\rho}_{x}) - A_0 \hat{p%
}_{xx} + u_{0}\rho_0 \hat{A}_{t} = 0.  \label{E39}
\end{equation}
Differentiating (\ref{E4}) gives: $p_{t}=\kappa \mbox{$\gamma$} \rho^{%
\mbox{$\gamma$} -1}\rho_{t}$, or $\hat{p}_{t} = \kappa \mbox{$\gamma$} \rho^{%
\mbox{$\gamma$} -1}_{0}\hat{\rho}_{t}$ upon linearizing at $\rho = \rho_0$.
Similarly, $\hat{p}_{x} = \kappa \mbox{$\gamma$} \rho^{\mbox{$\gamma$}
-1}_{0}\hat{\rho}_{x}$. With these relations, equation (\ref{E39}) becomes: 
\begin{eqnarray}
& & \mbox{$\Gamma$} (\partial_t +u_{0}\partial_x )^{2}
\hat{p} - \hat{p}_{xx} + {\frac{u_0 \rho_0 }{A_0%
}} \hat{A}_{t}  \nonumber \\
& + & A_{0}^{-1}\rho_0 (\partial_t +u_{0}\partial_x )^{2}
\hat{A} = 0,  \label{E40}
\end{eqnarray}
where: 
\[
\mbox{$\Gamma$} = {\frac{1}{\kappa \mbox{$\gamma$} \rho_{0}^{\mbox{$\gamma$}
-1}}} = {\frac{1}{c^{2}}}, 
\]
with $c$ being the speed of sound at air density $\rho_0$.

Applying the operator 
$A_0\mbox{$\Gamma$} (\partial_t +u_{0}\partial_x)^{2} 
- A_0 \partial_{xx}$ to both
sides of (\ref{E38}), we get: 
\[
A_0 [\Gamma \partial_{tt} + 2\Gamma u_{0}\partial_{xt} + (\Gamma u_{0}^{2}
-1)\partial_{xx}]
\cdot (\hat{A}_{tt} -\sigma \hat{A}_{xx}+ \mbox{$\alpha$} \hat{A}_{t} +
\beta \hat{A})
\]
\begin{equation}
= -4w\rho_0 [(\partial_{t}+u_{0}\partial_x)^{2}\hat{A} +
u_{0}\hat{A}_{t}].  \label{E41}
\end{equation}

Substituting the mode $\hat{A} = A_{m} 
e^{im^{\prime}x + \mbox{$\lambda$} t}$, 
$m^{\prime}={\frac{m\pi }{L}}$, we end up with the following algebraic
equation of degree four for $\mbox{$\lambda$}$: 
\[
[\mbox{$\Gamma$} \mbox{$\lambda$}^2 + 2\mbox{$\Gamma$} u_0 m^{\prime}%
\mbox{$\lambda$} i + (1 - \mbox{$\Gamma$} u_{0}^{2})m'^2] 
\cdot [\mbox{$\lambda$}^{2} + \sigma m'^2 +%
\mbox{$\alpha$} \mbox{$\lambda$} +\beta]
\]
\begin{equation}
= -4w\rho_0A_{0}^{-1}[\mbox{$\lambda$}^{2}+2 u_0 i m'
\mbox{$\lambda$} -u_0^{2}m'^2 ] - 
{\frac{4w u_{0}\rho_{0}\mbox{$\lambda$} }{A_0}},
\label{E42}
\end{equation}
or: 
\[
\mbox{$\Gamma$} \mbox{$\lambda$}^4 + (\mbox{$\alpha$} \mbox{$\Gamma$} + 2
\mbox{$\Gamma$} u_0 m^{\prime}i)
\mbox{$\lambda$}^3 + 
\]
\[
(\mbox{$\Gamma$} (\beta +\sigma m'^2) +2 \mbox{$\alpha$}
\mbox{$\Gamma$} u_0 m^{\prime}i  + 4
w\rho_0 A_0^{-1} + (1-\mbox{$\Gamma$} u_{0}^{2})m'^2 )
\mbox{$\lambda$}^2 
\]
\[
+ (2\mbox{$\Gamma$} u_0 m^{\prime}i 
(\beta +\sigma m'^2) -\mbox{$\alpha$} m'^2 (
\mbox{$\Gamma$} u_{0}^{2}-1) 
\]
\[
+ 
4w\rho_0A_{0}^{-1}(2u_0 i m')+{\frac{4w
u_0\rho_0 }{A_0}})\mbox{$\lambda$} 
\]
\begin{equation}
+ [ (\beta +\sigma m'^2)(1-\mbox{$\Gamma$} u_{0}^{2})m'^2
 + 4w\rho_0A_{0}^{-1}(- u_{0}^{2}m'^2)] = 0.  \label{E42a}
\end{equation}

\begin{prop}
 If 
\begin{equation}
\rho _{0}\,u_{0}^{2}>\mbox{$\alpha$}(\rho _{0}\,u_{0}+\mbox{$\Gamma$}\beta {%
\frac{u_{0}A_{0}}{4w}}),  \label{Osc}
\end{equation}
(\ref{E42a}) has a pair of pure imaginary solution $\lambda = \pm i\eta$, 
$\eta \not = 0$ being real, which implies the existence of a pair of 
oscillatory modes to the linearized system (\ref{E36})-(\ref{E38}) of 
the form $e^{\pm (im\pi x/L + i \eta t)}$, for real and nonzero 
numbers $m$ and $\eta$. The transition into oscillation is 
a Hopf bifurcation. 
\end{prop}

\noindent {\it Proof:} Let $\mbox{$\lambda$} = i \eta $ in
(\ref{E42a}), where $\eta $ is real. The real and imaginary parts give 
respectively: 
\begin{eqnarray}
\mbox{$\Gamma$} \eta^{4} & + & 2\mbox{$\Gamma$} u_{0}m^{\prime}\eta^{3} - [ %
\mbox{$\Gamma$} (\sigma m'^2 +\beta ) + m'^{2}(1-
\mbox{$\Gamma$} u_{0}^{2}) + {\frac{4 w\rho_0 }{A_0}}]\eta^{2}  \nonumber
\\
& + & [ -2 \mbox{$\Gamma$} u_{0}m^{\prime}(\beta +\sigma m'^{2}) -{%
\frac{8 w\rho_0u_0m^{\prime}}{A_0}}] \eta  \nonumber \\
& + & [m'^{2}(1-\mbox{$\Gamma$} u_{0}^{2})(\sigma
m'^{2}+\beta) - {\frac{4u_{0}^{2}m'^{2}w\rho_{0} }{A_0}}%
] = 0,  \label{E47}
\end{eqnarray}
and: 
\begin{equation}
-\mbox{$\alpha$} \mbox{$\Gamma$} \eta^{3} - 2\mbox{$\alpha$} \mbox{$\Gamma$}
u_{0}m^{\prime}\eta^{2} +m'^{2}\mbox{$\alpha$} (1-\mbox{$\Gamma$}
u_{0}^{2}) \eta +{\frac{4 w u_{0}\rho_{0} \eta }{A_{0}}} = 0.  \label{E48}
\end{equation}
For $\eta \not =0$, $\mbox{$\alpha$} \not = 0$, we have from (\ref{E48}): 
\[
\mbox{$\Gamma$} \eta^{2}+2\mbox{$\Gamma$} u_0 m^{\prime}\eta -
m'^{2}(1-\mbox{$\Gamma$} u_{0}^{2}) - {\frac{4 w u_0\rho_0 }{A_0 %
\mbox{$\alpha$}}} = 0, 
\]
so: 
\begin{equation}
\eta = -u_0 m^{\prime}\pm c \sqrt{ m'^{2} +4 w u_0\rho_0 /(A_0 %
\mbox{$\alpha$})}.  \label{E49}
\end{equation}
Now we regard the left hand side of (\ref{E47}) as a continuous function of $%
m^{\prime}$, call it $F(m^{\prime})$. For $|m^{\prime}| \gg 1$, $\eta \sim
(-u_0 \pm c)m^{\prime}$, direct calculation shows: 
\[
F(m^{\prime}) \sim - {\frac{4w\rho_0 }{\mbox{$\Gamma$} A_0}} m'^2
< 0. 
\]
While for $|m^{\prime}| \ll 1$, 
\[
\eta \sim \pm 2 w \sqrt{{\frac{u_0 \rho_0 }{\mbox{$\Gamma$} A_0 %
\mbox{$\alpha$}}}} + O(m^{\prime}), 
\]
and: 
\[
F(m^{\prime}) \sim {\frac{4 w u_0\rho_0 }{\mbox{$\Gamma$} A_0 %
\mbox{$\alpha$}}} \left ({\frac{4 w u_0\rho_0 }{A_0\mbox{$\alpha$}}} -%
\mbox{$\Gamma$} \beta - {\frac{4w\rho_0}{A_0}}\right ) > 0, 
\]
provided: 
\begin{equation}
{\frac{4 w u_0\rho_0 }{A_0}} > \mbox{$\alpha$} \left ({\frac{4 w\rho_0%
}{A_0}} +\mbox{$\Gamma$} \beta \right ),  \label{E50}
\end{equation}
holds, which is just (\ref{Osc}). 
Under (\ref{Osc}), $F(m)=0$ has a nonzero real solution, hence an oscillatory
mode solution exists to (\ref{E36})-(\ref{E38}). Finally, noticing 
that equations (\ref{E47})-(\ref{E48}) are invariant under the 
symmetry transform $(\eta, m') \ra (-\eta,-m')$, we conclude that 
the oscillatory modes exist as a pair, and oscillation appears as a 
Hopf bifurcation.

\begin{rmk}
Condition (\ref{Osc}) says that the fluid energy must be 
large enough to overcome the fold damping due to $\mbox{$\alpha$}$. 
Without the lower order term $A_t u/A$, the 
same  calculation would show that $F(m^{\prime}) =
-4w\rho_0 m'^2/ (\mbox{$\Gamma$} A_0) < 0$ for all $m^{\prime}$,
implying non-existence of oscillatory mode. 
Condition (\ref{E50}) is similar to the threshold pressure in Titze's model 
\cite{Ti} in the sense that minimum energy (analogous to minimum lung
pressure) is proportional to the fold damping coefficient and 
the prephonatory
half width $A_0$.
\end{rmk}

\begin{rmk}
There is another neutral yet nonoscillatory mode from 
(\ref{E47})-(\ref{E48}), namely, $\eta =0$, and: 
\begin{equation}
\sigma |m^{\prime}_0|^{2} = {\frac{4u_0^{2}w\rho_0 }{(1-M^2)A_0}} -\beta,
\label{E56}
\end{equation}
provided the right hand side expression is positive.
As we turn on $\mu >0$ but small, the oscillation mode 
in Proposition 2.1 will generically be
slightly perturbed and yet preserves its oscillatory nature. 
The calculations are tediuous and not shown here.
\end{rmk}

\section{Quasi-steady Approximation and Relations with the Titze Model}

\setcounter{equation}{0} 
The glottal flow is often regarded as nearly quasisteady and inviscid in 
the bulk \cite{Pel1}, and the Bernoulli's law is
 adopted as an approximation, \cite{Ishi}, \cite{Ti} among others. In
this approximation, the temporal variation of flow variables is considered
much slower than that of the fold motion. Now let us drop the time
derivatives, and viscous term in the flow equations to get: 
\begin{equation}
(\rho \, u \, A)_{x} =0,  \label{Q1}
\end{equation}
\begin{equation}
u\,u_{x} = -{\frac{p_{x}}{\rho}} +{\frac{A_{t}\,u }{A}},  \label{Q2}
\end{equation}
while keeping the fold dynamic equation and the equation of state the same.

Our goal is to derive a closed equation for the cross section area $A$.
Integrating (\ref{Q1}) and (\ref{Q2}) in $x$ using the equation of state
shows: 
\begin{eqnarray} 
& & \rho\, u \, A = Q_0,  \label{Q3} \\
& & u^{2}/2 + {\frac{\gamma \rho^{\gamma -1}}{\gamma -1}} = \int_{-L}^{x}{%
\frac{A_{t}\,u }{A}}\, dx\, + P_{0},  \label{Q4}
\end{eqnarray}
where $P_0$ and $Q_0$ are constants determined by the flow conditions at the
inlet $x=-L$. Note that without the integral term with ${\frac{A_{t}\,u }{A}}
$, (\ref{Q4}) becomes the standard Bernoulli's law.

Substituting (\ref{Q3}) into (\ref{Q4}) gives: 
\begin{equation}
{\frac{Q_0^2}{2\rho^2\,A^2}}+ {\frac{\gamma \rho^{\gamma -1}}{\gamma -1}} =
Q_0\, \int_{-L}^{x}{\frac{A_{t}\, }{\rho\, A^2}}\, dx\, + P_{0}.  \label{Q5}
\end{equation}
We would obtain a closed equation on $A$ if (\ref{Q5}) could be solved for $%
\rho$ in terms of $A$. 

Suppose that $A$ undergoes small vibration about a constant state $
A_0$ which satisfies: 
\[
{\frac{Q_0^2}{2\rho_0^2\,A_0^2}}+ {\frac{\gamma 
\rho_0^{\gamma -1}}{\gamma -1}}=P_0, 
\]
then (\ref{Q5}) can be solved for $\rho$ by perturbation. Letting $A=A_0 +%
\hat{A}$, $\rho=\rho_0+\hat{\rho}$, we find: 
\[
-{\frac{Q_0^{2}}{\rho_0^3\, A_0^2}}\hat{\rho} -{\frac{Q_0^2}{\rho_0^2A_0^2}} 
\hat{A}+\gamma\rho_0^{\gamma -2}\hat{\rho} -Q_0\int_{-L}^x\, {\frac{\hat{A}%
_t }{A_0^2\rho_0}} =0, 
\]
or: 
\begin{equation}
\hat{\rho}\cdot \left (\gamma\rho_0^{\gamma -2} - {\frac{Q_0^{2}}{\rho_0^3\,
A_0^2}}\right ) ={\frac{Q_0^3}{A_0^2\rho_0}}\hat{A}+ {\frac{Q_0}{A_0^2\rho_0}%
}\int_{-L}^x\, \hat{A}_t.  \label{Q6}
\end{equation}
If $\gamma\rho_0^{\gamma -2} > {\frac{Q_0^{2}}{\rho_0^3\, A_0^2}}$, or: 
\begin{equation}
\rho_0^{\gamma + 1} > {\frac{Q_0^2 }{\gamma \, A_0^2}},  \label{Q7}
\end{equation}
which means large pressure for given $Q_0$ and $A_0$, then: 
\begin{equation}
\hat{\rho} = {\frac{Q_0^2 }{\rho_0 A_0^2 }}\cdot {\frac{1 }{\gamma \rho_0^{%
\mbox{$\gamma$} -2} -{\frac{Q_0^{2}}{\rho_0^3\, A_0^2}} }} \left [
\int_{-L}^x \hat{A}_{t} + {\frac{Q_0}{\rho_0 A_0}}\hat{A}\right ].
\label{Q8}
\end{equation}

Now substituting (\ref{Q8}) into the fold dynamic $A$ equation, we obtain: 
\begin{equation}
\hat{A}_{tt}-\sigma \hat{A}_{xx}+\alpha \hat{A}_{t}+\beta \hat{A}%
=c_{aero}[\int_{-L}^{x}\hat{A}_{t}+{\frac{Q_{0}}{\rho _{0}A_{0}}}\hat{A}],
\label{Q9}
\end{equation}
where: 
\begin{equation}
c_{aero}={\frac{4w\kappa \gamma \rho _{0}^{\gamma -1}}{\gamma \rho _{0}^{%
\mbox{$\gamma$}-2}-{\frac{Q_{0}^{2}}{\rho _{0}^{3}\,A_{0}^{2}}}}}{\frac{%
Q_{0}^{2}}{\rho _{0}A_{0}^{2}}}>0.  \label{Q9a}
\end{equation}
Equation (\ref{Q9}) can be written as: 
\begin{equation}
\hat{A}_{tt}-\sigma \hat{A}_{xx}+(\alpha \hat{A}-c_{aero}\int_{-L}^{x}\hat{A}%
)_{t}+(\beta -c_{aero}Q_{0}\rho _{0}^{-1}A_{0}^{-1})\hat{A}=0.  \label{Q10}
\end{equation}
Equation (\ref{Q10}) is a linear wave equation with damping and pumping.
This is easy to see from the energy identity. Assuming that 
$\hat{A}_{x}(\pm L,t)=0$, we have: 
\begin{eqnarray}
&&{\frac{d}{dt}}\int_{-L}^{L}\,dx\,\left( {\frac{\sigma }{2}}\hat{A}_{x}^{2}+%
{\frac{1}{2}}\hat{A}_{t}^{2}+(\beta -c_{aero}Q_{0}\rho _{0}^{-1}A_{0}^{-1})%
\hat{A}^{2}/2\right) =  \nonumber \\
&&-\alpha \int_{-L}^{L}\,dx\,\hat{A}_{t}^{2}+c_{aero}(\int_{-L}^{L}\,dx\,%
\hat{A}_{t})^{2}/2.  \label{Q11}
\end{eqnarray}
We shall require that: 
\begin{equation}
\beta -c_{aero}Q_{0}\rho _{0}^{-1}A_{0}^{-1}>0,  \label{Q12}
\end{equation}
which is true if $\rho _{0}$ is sufficiently large for fixed $\beta $, 
$A_0$, and $
Q_{0}$. Then (\ref{Q11}) says that the rate of change of the total energy in
time depends on the balance of the natural fold damping (the negative term
with prefactor $\alpha $) and the energy input from the aerodynamic flow
(the positive term with prefactor $c_{aero}$).

We see from (\ref{Q11}) that spatially sinusoidal modes like
$e^{im' x}$ will render 
$\int_{-L}^{L} \hat{A}_t\, dx =0$, hence they do not sustain 
lossless temporall oscillations. However,
there are lossless oscillatory solutions with monotone spatial profiles.
To show a simple solution of this kind, 
extend the incoming flow uniformly to $-\infty $ and modify the energy
transfer term to $\int_{-\infty }^{x}\frac{\hat{A}_{t}u}{\hat{A}}$
in (\ref{Q10}). 
Let us keep the exit location still at $x=L$. Then seek $\varphi =e^{\nu
x}\psi \left( t\right)$, $\nu >0$, $\varphi _{x}=\hat{A}$, 
$\psi (t)$ satisfies:
\[
\nu \psi_{tt}+(\alpha \nu -c_{aero})\psi_{t}+\nu (\beta -c_{aero}Q_{0}\rho
_0^{-1}A_{0}^{-1}-\sigma \nu^{2})\psi =0.
\]

Since $\nu^{-1}$ is the length scale of spatial decay, $\nu^{-1}$ is of order  
$O(L).$  The critical condition for temporal oscillation is 
$\alpha \nu =c_{aero},$ to remove the damping, or 
\[
c_{aero}\geq \frac{\alpha }{L},
\]
or: 
\[
4w\kappa Q_{0}^{2}\geq \frac{\alpha A_{0}^{2}}{L}.
\]

\bigskip The temporal oscillation frequency is $\sqrt{\beta
-c_{aero}Q_{0}\rho _{0}^{-1}A_{0}^{-1}-\sigma \nu ^{2}},$ which is positive
for large $\rho _{0}$
and small $\sigma.$ Notice that the fold positions at $\pm L:A_0+\nu
e^{-\nu L}\psi (t)$, $A_{0}+\nu e^{\nu L}\psi (t),$
are not equal, they are above and below $A_0$ simultaneously. When they are
above $A_0,$
the fold is divergent (positive slope along the flow direction); and when
they are below $A_0$, the fold is convergent (negative
slope along the flow direction). However, the two end points are never
completely out of 
phase, i.e, when one is above and the other is below $A_{0}$.

The oscillatory solutions are of very different forms with or without 
making the quasi-steady approximation, which may 
underestimate the amount of the  
energy transfer from air flow onto the vocal fold. Nevertheless, 
the above spatially monotone solutions under the
quasi-steady approximation agree qualitatively with those of the Titze model 
\cite{Ti} and shows explicitly the effect of the energy transfer term ($
\int_{-L}^{x}{\frac{A_{t}u}{A}}$). This term is nonlocal and different from
the local term in Titze model \cite{Ti}, yet is of the same
origin and reflects the changing directions of the effective air driving
force on open and closing cycles, a key observation of the Titze model.

Both our model and the Titze model capture the positive energy feedback
from air flow into the fold motion, though they are not of the same form.
Our model has the advantage of a systematic treatment
and of characterizing the entire fold shape in the continuum. 

\vspace{.2 in}

\section{Acknowledgement}
The work of J.X. was partially supported by grants from NSF and 
Army Research Office. The work of J. M. Hyman was partially supported 
by Department of Energy. J.X. thanks Dr. I. Moise for helpful remarks.

\end{document}